\documentclass[a4paper,12pt]{article}
\usepackage[ascii]{inputenc}
\usepackage[T1]{fontenc}
\usepackage[english,english]{babel}
\usepackage{amsmath}
\usepackage{amssymb,amsfonts,textcomp}
\usepackage{color}
\usepackage{array}
\usepackage{supertabular}
\usepackage{hhline}
\usepackage{hyperref}
\newcounter{Text}

\begin{document}

\title{Rigid covariance as a natural extension of Painlev\'e--Gullstrand space-times: gravitational waves}
\author{Xavier Ja\'en\thanks{Dept. de F\'isica, Universitat Polit\`ecnica de Catalunya, Spain, e-mail address: xavier.jaen@upc.edu} and Alfred Molina\thanks{Dept. F\'{\i}sica Qu\`antica i Astrof\'{\i}sica, Institut de Ci\`encies del Cosmos (ICCUB), Universitat de Barcelona, Spain, e-mail address: alfred.molina@ub.edu}}

\maketitle

\begin{abstract}
The group of rigid motions is considered to guide the search for a natural system of space-time coordinates in General Relativity. This search leads us to a natural extension of the space-times that support Painlev\'{e}--Gullstrand synchronization. As an interesting example, here we describe a system of rigid coordinates for the cross mode of gravitational linear plane waves.
\end{abstract}
%
\section{Introduction}
At the beginning of his search for a theory of general relativity, Einstein's first steps were to search for a formulation of Special Relativity for non-inertial observers, using the Equivalence Principle to place inertial and gravitational forces on the same footing \cite{Einstein}.

The difficulties in carrying out that program led to Einstein introducing general covariance as a new principle and using it as a guide to clarify the way forward. Critics of such a strategy emerged, such as Kretschmann \cite{Kretschmann} and later Fock \cite{Fock}, who in general objected that since any theory can support a generally covariant formulation, the physical meaning of the principle was highly dubious \cite{Antoci}.

Essentially the situation is that General Relativity lacks of a dynamical invariance group equivalent to the group of rigid motions which characterize physical reference systems in Newtonian Mechanics.

Recently, some authors have suggested that may be advantageous to place some restriction to general covariance 
\cite{Ellis}. Such a restriction sometimes appears under the name of generalized isometries \cite{Bel}, \cite{Llosa}.

The concept of a rigid body and rigid motion arose naturally as an idealization of solid objects that surround us. In fact, from the point of view of both experience and physical theories, a perfectly rigid body cannot exist at the non-relativistic level, because it would imply the existence of infinite elastic modules. At the relativistic level, a new argument is added because the existence of a perfectly rigid body would imply instantaneous signal propagation between particles.

However, leaving aside its existence as a real substance, it is possible to conceive of rigid motion through some coherent construction; moreover, we are not necessarily interested in the existence of the substance that permits an implementation of the concept of rigid motion. Here, our interest is to study the compatibility between classical rigid motions and General Relativity. Some authors have argued that if we are able to implement the concept of rigid motion in the relativistic domain, we will be able to develop relativity to the same degree as we have developed Newtonian Mechanics; for instance, we will be able to develop a clear relativistic theory of elasticity.

The Painlev\'{e}--Gullstrand coordinate system \cite{Painleve,Gullstrand} is used to expand the Schwarzschild solution within its event horizon. Written in this coordinate system, the Schwarzschild metric is regular inside the horizon and it is singular only at $r=0$. Another interesting property is its spatial geometry: the surfaces $t=\rm{constant}$ are flat. This is what is known as Painlev\'{e}--Gullstrand synchronization. Such synchronization is interesting in the context of gravitational collapse due to the fact that we can go beyond the Schwarzschild radius. This kind of synchronization is increasingly present in the literature; for example, in the so-called analogue models of gravity \cite{Barcelo} or in relativistic hydrodynamics \cite{Ellis}. We will call \textit{Painlev\'{e}--Gullstrand space-times} those space-times that support a Painlev\'{e}--Gullstrand coordinate system.

In a series of previous papers \cite{Jaen1,Jaen2,Jaen3}, we established a close relationship between Painlev\'{e}--Gullstrand space-times and rigid motions. We showed how a 
significant set of space-times admit, as generalized isometries, the group of rigid motions. That set coincides with the set of Painlev\'{e}--Gullstrand space-times. The rigid  covariant formulation that we mention was built up by paying attention to some little-known  properties  of Newtonian Mechanics. This permitted  us to determine the physical meaning 
of the various potentials that arise quite well. We obtained a formulation of a set of space-times defined via five potentials which obey rigid covariant equations. 
This formulation does not cover all the space-times of General Relativity. Some of the space-times that are of particular interest to us, such as the Kerr space-time and the space-time that corresponds to gravitational waves, remain outside this formulation. 

Our aim in this paper is to establish whether rigid covariance can also support gravitational waves \cite{Misner}.

To this end,  we introduce a sixth potential while trying to maintain all the properties that are characteristic of the rigid covariant formulation developed so far. In particular, we are looking for a covariant formulation under a group of transformations that allows us to characterize the space-time metric by means of the six potentials.  The group of rigid motions is a reasonable candidate to play this role, as it already does in Newtonian Mechanics, and there is no \textit{a priori} reason not to consider it. We believe that studying the possibility of formulating General Relativity, or a significant portion thereof, in a way that is covariant under the group of rigid motions, using essentially six potentials, is work that is important in itself, beyond the interpretations that may arise. We prove that this is indeed possible and as a result, we will obtain a rigid system of coordinates for gravitational linear plane waves.

For the sake of simplicity, in this paper we fix the value of the cosmological potential $H$ \cite{Jaen2} to $H = 1$. Our work does not depend at all on this condition and it can easily be implemented if $H \neq 1 $.

So, in section\S\ref{sec_2}, we show how the usual Painlev\'{e}--Gullstrand space-times can be understood as  rigid covariant space-times. We then review their main properties from this new perspective. In section\S\ref{sec_3}, we study an extension of the 
Painlev\'{e}--Gullstrand space-times  which maintain the rigid covariance together with most of the properties studied in the previous section. Then, section\S\ref{sec_4}  is devoted to finding a rigid system of coordinates for a given space-time in arbitrary  coordinates.
Finally, in section\S\ref{sec_5}, we apply the equations derived in the previous sections to finding a rigid system of coordinates for a gravitational linear plane wave. 
%
\section{The rigid covariant formulation of Pain\-le\-v\'e--Gullstrand space-times}\label{sec_2}

We define Painlev\'e--Gullstrand space-times as those that admit of space-time coordinates (known by the same name) in such a way that the metric allows the possibility of flat space slicing.

The metric of a Painlev\'e--Gullstrand space-time  can always be written using four potentials,  $\Phi,K_{i}$, in the form:\footnote{Throughout the paper we will use the following notation:
Latin indices  $i,j,k=1,2,3$; $dx\,dy=\frac{1}{2}(dx\otimes
dy+dy\otimes dx)$; $T_{(i}Q_{j)}=\frac{1}{2}(T_{i}Q_{j}+T_{j}Q_{i})$; $\delta _{ij}$ is the three-dimensional identity;
$f_{,i}=\frac{\partial f}{\partial x^{i}}$, where  $x^{i}$ are the space coordinates; $f_{,\lambda }=\frac{\partial
f}{\partial \lambda }$;  $\bar{d}$ is the restriction of the differential to  $d\lambda =0$, i.e. $\bar{d} f(\vec x,\lambda )= \frac{\partial f}{\partial x^{i}} dx^i$.
}
\begin{equation}\label{eq_1_metric}
ds^{2}=-\Phi^{2}d\lambda
^{2}+2K_{i}dx^{i}d\lambda +\delta
_{ij}dx^{i}dx^{j}
\end{equation}
We define the potential $\tau$ and ${\vec{v}}$ according to 
\begin{eqnarray}\label{eq_2}
&& \Phi^{2}=-\delta_{ij}v^{i}v^{j}+c^{2}\left( \tau_{,\lambda }^2-(\tau_{,i}v^{i})^{2} \right)\nonumber\\
&& K_{i}=-\delta_{ij} v^{j}-c^{2}\left(\tau _{,\lambda }+\tau_{,j}v^{j}\right)\tau _{,i}
\end{eqnarray}
 One can see \cite{Jaen3} that  $\tau$ is any solution of the action in the Hamilton--Jacobi equation associated with the metric (\ref{eq_1_metric}). That is, $\tau$ is any solution of: 
\begin{equation}\label{eq_4}
\partial _{\lambda }\tau+H(\vec{x},\vec{p}=\vec{\nabla }\tau,\lambda
)=0
\end{equation} 
with
\begin{equation}\label{eq_3}
H(\vec{x},\vec{p},\lambda )=-{\vec{K}}\cdot
\vec{p}-\sqrt{[1+c^{2}{\vec{p}}^{2}]\left[\left(\frac{K}{c}\right)^{2}+\frac{\Phi^{2}}{c^{2}}\right]}
\end{equation}

For each solution,  $\tau$, of (\ref{eq_4}), the corresponding potential 
${\vec{v}}$ is:
\begin{equation}\label{eq_5}
{\vec{v}}=\frac{\partial H}{\partial
\vec{p}}(\vec{x},\vec{p}=\vec{\nabla }\tau,\lambda )
\end{equation}
In terms of the potentials  $\tau $  and  ${\vec{v}}$, the metric
 (\ref{eq_1_metric}) can be written as:
\begin{equation}\label{eq_6}
\mathit{ds}^{2}=-c^{2}d\tau^{2}+c^{2}[\vec{\nabla }\tau\cdot
(d\vec{x}-{\vec{v}}d\lambda )]^{2}+(d\vec{x}-{\vec{v}}d\lambda
)^{2}
\end{equation}
which has the following properties:
\begin{enumerate}
\item Newtonian limit:

If we have that  $\tau=\lambda +\frac{f(\vec{x},\lambda )}{c^{2}}$, the Newtonian non-relativistic limit can be obtained as  $c\to \infty $ without any consideration regarding weak fields \cite{Jaen1}.
\item Rigid motion covariance or \textit{rigid motion generalized isometry}\cite{Bel,Llosa}:

This is a property of space-time that is not apparent from the perspective of a metric in the form of (\ref{eq_1_metric}), but instead in the form (\ref{eq_6}) it becomes quite natural. Under rigid motions transformations
\begin{eqnarray}\label{eq_7}
&& \lambda =\lambda' \nonumber
\\ && \vec{x}\equiv x^{i}\vec{e}_{i} = \vec{X}(\lambda
)+\vec{x}{}'=X^{i}(\lambda
)\vec{e}_{i}+x{}'^{i}\vec{e}_{i}{}'= \nonumber
\\ && \hspace*{1em} \left(X^{i}(\lambda
)+x{}'^{k}R_{k}^{i}(\lambda )\right)\vec{e}_{i},
\end{eqnarray}
where  $R_{k}^{i}(\lambda )$ is an orthogonal matrix, (\ref{eq_6}) is shape invariant.
To be more specific, the metric 
becomes:
\begin{equation}
\mathit{ds}^{2}=-c^{2}d\tau'^{2}+c^{2}[\vec{\nabla }\tau'
\cdot (d\vec{x}{}'-\vec{v}{}'d\lambda
)]^{2}+(d{\vec{x}}{}'-{\vec{v}}{}'d\lambda )^{2}
\end{equation}
with  $\tau'({\vec{x}}{}',\lambda )=\tau(\vec{x},\lambda )$ and $\vec{v}{}'(\vec{x}{}',\lambda)=\vec{v}(\vec{x},\lambda )-\vec{v}_{0}(\vec{x},\lambda )$; and where ${\vec{v}}_{0}(\vec{x},\lambda )$ is the field associated with the rigid trajectories (7), i.e.:
$$
{\vec{v}}_{0}(\vec{x},\lambda
)=\dot{{\vec{X}}}(\lambda )+\vec{\Omega }(\lambda )\times
[\vec{x}-\vec{X}(\lambda )],
$$
where  $\vec{\Omega }(\lambda)=\frac{1}{2}\sum _{j}{R_{j}}^{k}(\lambda ){{\dot{R}}_{j}}^{m}(\lambda){\vec{e}}_{k}\times {\vec{e}}_{m}$ and  $\times$ stands for the usual cross product.

We will say that (\ref{eq_6}) is the manifestly rigid covariant form of the metric.

\item Physical meaning of the potentials $\tau$ and  $\vec{v}$:

By construction, expressions (\ref{eq_4}) and (\ref{eq_5}), and  from the  metric (\ref{eq_6}), we see that the field $U=\partial_{\lambda}+\vec{v} \cdot \partial_{\vec{x}}$   is geodesic with proper time  $\tau$ \cite{Jaen3}. 
\item Gauge invariance:
  
We saw this invariance earlier when we defined the potentials $\tau$ and  ${\vec{v}}$ in (\ref{eq_2}). In the present context, if we start with the potentials  $\tau$ and ${\vec{v}}$, which using (\ref{eq_2}) \ give  $\Phi $ and  $K_{i}$, then any solution  $\tau ^{\text{*}}$ and ${\vec{v}}^{\text{*}}$ of Equations (4) and (5) will give, again using (2), the same potentials:  $\Phi$ and  $K_{i}$. This gauge invariance has a clear meaning because of the physical meaning of the potentials  $\tau$ and  ${\vec{v}}$. 
\item Painlev\'e--Gullstrand synchronization:

The slicing $\lambda =\text{constant}$ is flat; i.e.: $ds^{2}|_{d\lambda =0}=d{\vec{x}}^{2}$.
\end{enumerate}
By tending towards the limit $c\to \infty $, we can see how the meaning of the potential  $\vec{v}$,  the gauge invariance and the rigid covariance persist at a Newtonian level. As shown in \cite{Jaen1}--\cite{Jaen3}, regardless of General Relativity, a Newtonian theory of gravitation can be formulated starting from a potential, ${\vec{v}}$, in such a way that it unifies the inertial and gravitational fields in a set of equations that are shape invariant under rigid motion transformations. In this theory, the integral trajectories of ${\vec{v}}$ are solutions of the equation of motion for test particles. In fact, first we found the properties \textbf{2--4}  at a Newtonian level and later we used them as a guide to define the metric (\ref{eq_6}), as is explained in \cite{Jaen3}.

\section{The rigid motion covariant form of the metric}\label{sec_3}
In this section we introduce a new potential to generalize the metric (\ref{eq_6}). It should be borne in mind that, for simplicity, we have set the value of $H = 1$. The new potential that we aim to introduce has nothing to do with the potential $H$.  The price we will pay is such that we will lose the flat space slicing property. What we will see is that a new potential can be introduced while maintaining properties \textbf{1--4}. We start by considering the metric:
\begin{equation}
ds^{2}=-c^{2}d{\tau}^{2}+c^{2}(\tau
_{,i}(dx^{i}-v^{i}d\lambda ))^{2}+\gamma
_{ij}(dx^{i}-{v}^{i}d\lambda
)(dx^{j}-{v}^{j}d\lambda )
\end{equation}
where  $\gamma _{ij}$ is not specified and may functionally depend on the potentials  $\tau,{\vec{v}}$ and on a new potential denoted by $\sigma$. Following the same steps as in \cite{Jaen3}, we can see that if $\gamma_{ij}$ does not depend on $\tau$ and ${\vec{v}}$,  and if furthermore $\sigma$ is gauge invariant, then $\gamma_{ij}$ will also be gauge invariant. Then Equations (\ref{eq_1_metric}--\ref{eq_5}) will only be modified by the fact that, instead of using $\delta_{ij}$, we use  $\gamma_{ij}$. In order to maintain property {\textbf{2}}, we will require that  $\gamma_{ij}( dx^{i}-v^{i}d\lambda
)(dx^{j}-{v}^{j}d\lambda )$ be rigid covariant. Under these conditions, we have a couple of candidates: $\gamma_{ij}\equiv\delta _{ij}+\epsilon \sigma _{,i}\, \sigma_{,j}$ where $\epsilon =\pm 1$. 

In a coordinate system  $\{\lambda , x^{i}\}$, that we call \textit{rigid Euclidean}, the family of metrics:
\begin{equation}\label{eq_11}
ds^{2}=-\Phi^{2}d\lambda^{2}+2K_{i}dx^{i}d\lambda +\left(\delta_{ij}+\epsilon\sigma_{,i}\sigma_{,j}\right)dx^{i}dx^{j}
\end{equation}
have properties \textbf{1--4}  with the following modifications:

\textbf{1.} If  $\sigma =\frac{1}{c}s$, the Newtonian non-relativistic limit can still  be obtained as  $c\to \infty $ without any considerations regarding weak fields.

\textbf{2.} The space slicing  $\lambda =\text{constant}$ becomes a minimum modification of the flat case:
$ds^{2}|_{d\lambda=0}=\bar{d}{\vec{x}}^{2}+\epsilon(\bar{d}\sigma)^{2}.$ 

The expression of the metric (\ref{eq_11}) is the basis of rigid General Relativity.  $\epsilon =\pm 1$  with the sign to be determined. The five potentials of rigid General Relativity are: $\Phi,K_{i}$ and $\sigma$ (six, if we also consider the cosmological potential $H$). We can express  $\Phi, K_{i}$  in terms of the potentials  $\tau,v^{i}$  and  $\sigma$. We will have a gauge freedom in the choice of  $\tau,v^{i}$. With  $\gamma_{ij}\equiv \delta_{ij}+\epsilon \sigma _{,i}\sigma_{,j}$, the relationship between the potentials  $\Phi ,K_{i}$  and  $\tau,v^{i}$ and  $\sigma$  is now:
\begin{eqnarray}
&&\Phi^{2}=-\gamma_{ij}v^{i}v^{j}+c^{2}\left(\tau _{,\lambda }^{2}-(\tau
_{,i}v^{i})^{2}\right) \nonumber \\ && K_{i}=-\gamma
_{ij}v^{j}-c^{2}\left(\tau _{,\lambda }+\tau
_{,j}v^{j}\right)\tau _{,i}
\end{eqnarray}
i.e., the same expression as in (\ref{eq_2}) but using $\gamma_{ij}$ instead of $\delta_{ij}$.

The metric (\ref{eq_11}) in terms of these potentials becomes:
\begin{equation}\label{eq_13}
ds^{2}=-c^{2}d\tau^{2}+c^{2}(\tau
_{,i}(dx^{i}-v^{i}d\lambda ))^{2}+\left(\delta
_{ij}+\epsilon \sigma _{,i}\sigma
_{,j}\right)(dx^{i}-v^{i}d\lambda
)(dx^{j}-v^{j}d\lambda )
\end{equation}
which is the manifestly rigid covariant form of General Relativity.

Note that if we also consider the cosmological potential, $H$, the only change we need to implement in (\ref{eq_13}), and throughout the entire paper, is to replace the flat Euclidean metric $\delta_{ij}$ by $H^{-2}\delta_{ij}$. The properties \textbf{1-4} studied in \S\ref{sec_2}   will be maintained; and in this case, property \textbf{5}, the slicing $\lambda =\text{constant}$, will be $ds^{2}|_{d\lambda=0}=H^{-2} \bar{d}{\vec{x}}^{2}+\epsilon(\bar{d}\sigma)^{2}$.
%
\section{Moving from general covariance to  rigid covariance}\label{sec_4}
Given a metric  written in unspecified  space-time coordinates $\{T,X^{i}\}$:
\begin{equation}\label{eq_14}
ds^{2}=-\Phi
^{2}dT^{2}+2K_{i}dX^{i}dT+\gamma
_{ij}dX^{i}dX^{j}
\end{equation}
i.e., given the ten known coefficients $\{\Phi ,K_{i},\gamma _{ij}\}$, we aim to find the same metric but written in a rigid Euclidean coordinate system  $\{\lambda ,x^{i}\}$. The form (\ref{eq_14})  of the metric  is generally covariant: it contains ten potentials. We want to write the same metric in the rigid covariant form.

First we perform a time transformation  $T=T(\lambda ,X^{i})$ so that (\ref{eq_14}) becomes:
\begin{equation}\label{eq_15}
ds^{2}=-\Phi ^{2}T_{,\lambda }^{2}d\lambda ^{2}+2T_{,\lambda
}[K_{i}-\Phi ^{2}T_{,i}]dX^{i}d\lambda +[\gamma
_{ij}+2K_{i}T_{,j}-\Phi ^{2}T_{,i}T_{,j}]dX^{i}dX^{j}
\end{equation}
We want the space components of the metric (\ref{eq_15}) to take the form:
\begin{equation} \label{eq_16}
\gamma _{ij}+2K_{\text{(}i}T_{,j\text{)}}-\Phi
^{2}T_{,i}T_{,j}=\Delta _{ij}+\epsilon \sigma _{,i}\sigma
_{,j}
\end{equation}
where $\Delta _{ij}$ must be a three-dimensional flat metric. Solving for $\Delta _{ij}$:
\begin{equation} \label{eq_17}
\Delta _{ij}=\gamma _{ij}+2K_{\text{(}i}T_{,j\text{)}}-\Phi
^{2}T_{,i}T_{,j}-\epsilon \sigma _{,i}\sigma _{,j}
\end{equation}
To determine  $T(\lambda ,X^{i})$   and  $\sigma(\lambda ,X^{i})$, we require
$\Delta _{ij}$  be flat. Regardless of the nature of the  $X^{i}$ coordinates, this condition  can be expressed as:
\begin{equation}\label{eq_18}
Ricci_{(3)}(\Delta _{ij})=0
\end{equation}
Since the generalization to $H\neq 1$ is trivial, we can assert that if the corresponding Equation (\ref{eq_18}), in accordance with the comments at the end of section \S\ref{sec_3} by replacing $\delta_{ij}\longrightarrow H^{-2}\delta_{ij}$ in $\Delta _{ij}$ with the unknowns $T(\lambda ,X^{i})$, $\sigma(\lambda ,X^{i})$   and $H(\lambda ,X^{i})$, has a solution for some functions $T$, $\sigma$  and $H$, then rigid covariance using the six essential potentials will be locally equivalent to general covariance.

Once we have found  $T$  and  $\sigma$ from (\ref{eq_18}), we can find a rigid Euclidean coordinate system  $x^{i}$. Generally,  $\Delta _{ij}$, despite being flat, will not have the Euclidean (canonical) form  $\delta _{ij}$. Therefore, we can perform a change 
$X^{i}=X^{i}(\lambda ,x^{k})$,  so:
\begin{equation} \label{eq_19}
\Delta _{mn}X_{,i}^{m}X_{,j}^{n}=\delta _{ij}
\end{equation}
Note that when performing the change  $X^{i}=X^{i}(\lambda ,x^{k})$  on $\Delta _{ij}$, we only change the space coordinates  $X^{i}$. The time  $\lambda $ in the expression  $X^{i}=X^{i}(\lambda ,x^{k})$ is only a parameter. The change is possible 
if (\ref{eq_18}) can be solved for $T$ and $\sigma$.

Finally the change $\{\vec{X},T\}\to \{\vec{x},\lambda\}$, together with composition of the changes that we have found,  $\left\{X^{i}=X^{i}(\lambda ,x^{k}),\, T=T(\lambda ,\ X^{i}(\lambda,x^{k}))\right\}$, will transform the metric (\ref{eq_14}), written in the coordinates 
$\{\vec{X},T\}$, into the form (\ref{eq_11}) in rigid Euclidean coordinates  
$\{\vec{x},\lambda\}$.
%
\section{Gravitational linear plane waves}\label{sec_5}
In this section, we want to find a  rigid covariant form for a gravitational linear plane wave. In the coordinates $\{\vec{X}=(X,Y,Z),T\}$, consider the cross mode of a linear plane wave \cite{Misner} $h_{\times}=h$  (we take  $c=1$)
\begin{equation}\label{eq_20}
ds^{2}=-dT^{2}+{d\vec{X}}^{2}+2h\;\varepsilon
^{2}dX\,dY
\end{equation}
with  $h=h(Z-T)$. (\ref{eq_20}) is everywhere a solution of  $Ricci=0$ with  $Riemann\neq 0$   up to order  $\varepsilon ^{2}$. In what follows, we will work up to order  $\varepsilon ^{2}$. We note that the  coordinate system $\{\vec{X},T\}$ are adapted geodesic coordinates; i.e., the lines  $\vec{X}$ constant are geodesics. Comparing (\ref{eq_20}) and  (\ref{eq_14}), we have:
 \begin{equation} \label{eq_21}
\Phi ^{2}=1\ ;\ K_{i}=0\ ;\ \gamma ={d\vec{X}}^{2}+2h\varepsilon
^{2}\mathit{dX}\mathit{dY}
\end{equation}
The expression (\ref{eq_17}) for the metric  $\Delta _{ij}$  is:
\begin{equation} \label{eq_22}
\Delta _{ij}=(\gamma _{ij}-T_{,i}T_{,j}-\epsilon \sigma _{,i}\sigma
_{,j})dX^{i}dX^{j}
\end{equation}
Performing the time transformation  $T=\lambda +\varepsilon(XT_{(x)}+YT_{(y)})$ and  choosing $\sigma=\varepsilon (X\sigma _{(x)}+Y\sigma _{(y)})$, where $\{T_{(x)},T_{(y)},\sigma _{(x)},\sigma _{(y)}\}$ are functions depending on  $Z-\lambda $, the condition  $Ricci_{(3)}(\Delta _{ij})=0$ up to order $\varepsilon ^{2}$  demands   $\epsilon =-1$ and:\footnote{In this section we will use a prime, $f'$, to indicate the derivation of a function with respect to its argument.}
\begin{equation}\label{eq_23}
\sigma _{(x)}'{}^{2}=T'_{(x)}{}^{2}; \,\sigma
_{(y)}'^{2}=T'_{(y)}{}^{2}\ ;\ 2\sigma _{(x)}'\sigma
_{(y)}'-2T_{(x)}'T_{(y)}'+h''=0
\end{equation}
If we take:
\begin{equation}\label{eq_24}
\sigma'_{(x)}=T'_{(x)};\ \sigma'
_{(y)}=-T'_{(y)}
\end{equation}
which fulfil the first two conditions of (\ref{eq_23}), the third condition of (\ref{eq_23}) becomes:
\begin{equation}\label{eq_25}
4T'_{(x)}T'_{(y)}=h''
\end{equation}
which can always be fulfilled for any function $h$.

To complete the work, we must find a system of rigid Euclidean coordinates. We can solve  $\Delta _{ij}dX^{i}dX^{j}=\delta
_{ij}dx^{i}dx^{j}$  for  a coordinate change $\vec{X}=\{X,Y,Z\}\to \vec{x}=\{x,y,z\}$. This change depends on $\lambda $, which in the space  $\vec{X}$ acts as a parameter. Linking the two transformations,  $\{X,Y,Z,T\}\to \{x,y,z,\lambda \}$ and up to the order  $\varepsilon ^{2}$:
\begin{equation}\label{eq_26}
\begin{matrix}
X=x+\varepsilon^{2}y\left[T_{(x)}T_{(y)}-{\displaystyle \frac{h}{2}}+{\displaystyle \int}(T_{(x)}T'_{(y)}-T_{(y)}T'_{(x)})dz\right]
\hfill\null\\[2ex]
Y=y+\varepsilon ^{2}x\left[T_{(x)}T_{(y)}-{\displaystyle\frac{h}{2}}-{\displaystyle\int}
(T_{(x)}T'_{(y)}-T_{(y)}T'_{(x)})dz\right]\hfill\null\\[2ex]
Z=z+{\displaystyle\frac{\varepsilon ^{2}}{2}}xyh'\hfill\null \\[2ex]
T=\lambda+\varepsilon \left(xT_{(x)}+yT_{(y)}\right)\hfill\null 
\end{matrix}
\end{equation}
where $T_{(x)},T_{(y)}$ and  $h$ can be considered functions on $z-\lambda $, and we should recall that it is necessary to fulfil
$4T'_{(x)}T'_{(y)}=h''$. If we perform the change (\ref{eq_26}) on the metric ({\ref{eq_20}), we obtain a rigid covariant expression for this metric which agrees with (\ref{eq_11}).
%
\subsection{The monochromatic linear plane wave}
A particularly interesting case is that of the cross mode of a monochromatic linear plane wave, with frequency $\omega $. This corresponds to considering (\ref{eq_20}) with:
\begin{equation} \label{eq_28}
\varepsilon ^{2}\;h(Z,T)=A^{2}\sin (\omega (Z-T))
\end{equation}
i.e. $\varepsilon =A$ and  $h=\sin [\omega (Z-\lambda )]$ . As a solution of (\ref{eq_25}), we choose:
\begin{equation}\label{eq_29}
T_{(x)}=\sqrt{2}\sin [\frac{\omega }{2}(Z-\lambda
)]\ ;\ T_{(y)}=\sqrt{2}\cos [\frac{\omega }{2}(Z-\lambda )]
\end{equation}
Using (\ref{eq_26}) and working always up to order $A^{2}$, we obtain the coordinate change:
\begin{equation}
\begin{matrix}
X=x+yA^{2}\left\{{\displaystyle\frac{1}{2}}\sin [\omega (z-\lambda )]-\omega(z-\lambda )+f_{x}(\lambda )\right\}\hfill\null \\[2ex]
Y=y+xA^{2}\left\{{\displaystyle\frac{1}{2}}\sin\omega (z-\lambda )]+\omega (z-\lambda )+f_{y}(\lambda )\right\}\hfill\null\\[2ex] 
Z=z+xy{\displaystyle\frac{1}{2}}\omega A^{2}\cos [\omega (z-\lambda )]\hfill\null\\[2ex]
T=\lambda +\sqrt{2}A\left\{x\sin\left[{\displaystyle\frac{\omega }{2}}(z-\lambda))\right]+y\cos\left[{\displaystyle\frac{\omega }{2}}(z-\lambda )\right]\right\}\hfill\null
\end{matrix}
\end{equation}
where we have included the two  arbitrary functions of $\lambda $, $f_{x}(\lambda )$  and  $f_{y}(\lambda )$, as a consequence of the pair of integrals on $z$ appearing in (\ref{eq_26}). The inverse change is:
\begin{equation}\label{eq_30}
\begin{matrix}
x=X-YA^{2}\left\{ {\displaystyle\frac{1}{2}}\sin [\omega (Z-\lambda )]-\omega(Z-\lambda )+f_{x}(\lambda )\right\}\hfill\null \\[2ex]
y=Y-XA^{2}\left\{ {\displaystyle\frac{1}{2}}\sin
[\omega (Z-\lambda )]+\omega (Z-\lambda )+f_{y}(\lambda )\right\}\hfill\null\\[2ex]
z=Z-XY{\displaystyle\frac{1}{2}}\omega A^{2}\cos [\omega (Z-\lambda )]\hfill\null \\[2ex]
T=\lambda +\sqrt{2}A\left\{X\sin\left[{\displaystyle\frac{\omega }{2}}(Z-\lambda)\right]+Y\cos\left[{\displaystyle\frac{\omega }{2}}(Z-\lambda )\right]\right\}\hfill\null
\end{matrix}
\end{equation}
Since $\vec{X}$ are adapted geodesic coordinates, we can interpret (\ref{eq_30}) as  geodesic trajectories $\vec{x}(\lambda ;\vec{X})$ with proper time  $T(\lambda ;\vec{X})$ and $\vec{X}$ playing the role of the initial conditions.

We can also find, in  rigid coordinates, the geodesic velocity field or potential  ${\vec{v}}$, ${\vec{v}}(\vec{x},\lambda)=\left.\frac{\partial \vec{x}(\lambda;\vec{X})}{\partial \lambda }\right|_{\vec{X}\rightarrow
\vec{x}}$, and the corresponding proper time field or potential} $\tau$, $\tau(\vec{x},\lambda )=\left.T(\lambda;\vec{X})\right|_{\vec{X}\rightarrow \vec{x}}$, which, together with $\sigma(\vec{x},\lambda )=\left.\varepsilon(X\sigma_{(x)}+Y\sigma _{(y)})\right|_{\vec{X}\rightarrow \vec{x}}$  and (\ref{eq_24}), characterize the space-time of the wave (\ref{eq_20}).

For  $\lambda =0$,  from (\ref{eq_30}) we can define $x_0=x(\lambda=0,\vec X)$, $y_0=y(\lambda=0,\vec X)$ and $z_0=z(\lambda=0,\vec X)$ and if we perform the transformation  $\vec{X}=\{X,Y,Z\}\to\vec{y}=\{x_{0,}y_{0,}z_{0}\}$ on (\ref{eq_30}), we have:
\begin{equation}\label{eq_32}
\begin{matrix}
x=x_{0}-y_{0}A^{2}\left\{{\displaystyle\frac{1}{2}}\left(\sin [\omega
(z_{0}-\lambda )]-\sin [\omega z_{0}]\right)-\omega \lambda
+f_{x}(\lambda )-f_{x}(0)\right\}\hfill\null\\[2ex]
y=y_{0}-x_{0}A^{2}\left\{{\displaystyle\frac{1}{2}}\left(\sin [\omega (z_{0}-\lambda
)]-\sin [\omega z_{0}]\right)+\omega \lambda +f_{y}(\lambda
)-f_{y}(0)\right\}\hfill\null \\[2ex]
z=z_{0}-x_{0}y_{0}{\displaystyle\frac{1}{2}}\omega A^{2}\cos
[\omega (z_{0}-\lambda )]\hfill\null \\[2ex]
T=\lambda+\sqrt{2}A\left\{x_{0}\sin\left[{\displaystyle\frac{\omega}{2}}(z_{0}-\lambda
)\right]+y_{0}\cos\left[{\displaystyle\frac{\omega }{2}}(z_{0}-\lambda )\right]\right\}\hfill\null
\end{matrix}
\end{equation}
The geodesic corresponding to the initial conditions $\{x_{0,}y_{0,}z_{0}\}=\{0,0,0\}$  is  $\{x,y,z\}=\{0,0,0\}$.
Choosing  $f_{x}(\lambda )=-f_{y}(\lambda )=\omega \lambda $  (\ref{eq_32}) becomes:
\begin{equation}\label{eq_33}
\begin{matrix}
x=x_{0}-y_{0}A^{2}\left\{{\displaystyle\frac{1}{2}}\left(\sin [\omega
(z_{0}-\lambda )]-\sin[\omega z_{0}]\right)\right\}\hfill\null\\[2ex]
y=y_{0}-x_{0}A^{2}\left\{{\displaystyle\frac{1}{2}}\left(\sin [\omega (z_{0}-\lambda
)]-\sin [\omega z_{0}]\right)\right\}\hfill\null\\[2ex]
z=z_{0}-x_{0}y_{0}{\displaystyle\frac{1}{2}}\omega A^{2}\cos [\omega (z_{0}-\lambda
)]\hfill\null \\[2ex]
T=\lambda +\sqrt{2}A\left\{x_{0}\sin\left[{\displaystyle\frac{\omega
}{2}}(z_{0}-\lambda )\right]+y_{0}\cos\left[{\displaystyle\frac{\omega }{2}}(z_{0}-\lambda
)\right]\right\}\hfill\null 
\end{matrix}
\end{equation}
which, to first order in the coordinates near  $x_{0}=y_{0}=z_{0}=0$ becomes:
\begin{equation}\label{eq_34}
\begin{matrix}
x=x_{0}+y_{0}A^{2}{\displaystyle\frac{1}{2}}\sin[\omega \lambda]\hfill\null \\[2ex]
y=y_{0}+x_{0}A^{2}{\displaystyle\frac{\omega}{2}}\lambda{\displaystyle\frac{1}{2}}\sin [\omega \lambda
]\hfill\null \\[2ex]
z=z_{0}\hfill\null \end{matrix}
\end{equation}
and
\begin{equation}\label{eq_35}
T=\lambda+\sqrt{2}A\left((y_0\cos\left[{\displaystyle\frac{\omega}{2}}\lambda\right]-x_0\sin\left[{\displaystyle\frac{\omega}{2}}\lambda\right]\right)
\end{equation}
This coincides with the usual result \cite{Misner}. We note that, up to the order in which we work, we can replace 
$\lambda=T$  in  (\ref{eq_34}). In fact, $\lambda $ is the proper time of the geodesic  $x_{0}=y_{0}=z_{0}=0$.
%
\section{Conclusions}
In this paper we have tried to advance the review of some aspects of the foundations of General Relativity that we began in three recently published papers \cite{Jaen1,Jaen2,Jaen3}. In \cite{Jaen3} we identified up to five metric potentials with physical meaning. There, we saw how, using these potentials, we could express the metric of a significant set of space-times in a rigid Euclidean coordinate system. However, we realized that the Kerr space-time and those related to gravitational waves remain outside that set. 

In the present paper we have conveniently introduced a sixth potential, $\sigma$, completing a minimal set of independent potentials to try to cover, locally by using a rigid Euclidean coordinate system, the whole of General Relativity. As a significant example, we have written the space-times of a gravitational linear plane wave in a rigid Euclidean coordinate system.

It is important to note that in doing so we have had no need to use any kind of Fermi coordinates \cite{Manasse}. That is, our rigid Euclidean coordinate system is an exact concept in General Relativity and does not arise as a consequence of 
any kind of approximation  process. The only approximation we have made is related exclusively to the fact that in section \S\ref{sec_5} we are working with linear waves.

This does not mean that our proposal is free from difficulties. We have found a rigid Euclidean coordinate system for gravitational linear plane waves, but we were not been able to guarantee its existence before the calculation neither do we have a well-defined uniqueness criterion that would guarantee a unique rigid Euclidean coordinate system, except for changes related to the physical observer, as is the case in Newtonian Mechanics. Regarding gravitational waves, what we have proven is that we can find a rigid Euclidean coordinate system  from which, by using the usual approximation, we obtain the known results. But we do not know the meaning of the expressions we found without using the same kind of approximation that people usually do when studying gravitational waves, which is none other than the use of Fermi coordinates. To take advantage of the rigid coordinates found, we think it will require a little more work along the lines set out in the following paragraphs.

Given an arbitrary space-time, the existence of a rigid Euclidean coordinate system is guaranteed if we can prove that Equation (\ref{eq_18}), taking into account the cosmological potential $H$, always has a solution for some functions $T$, $\sigma$ and $H$. This is an open problem. As we state above, if we are able to prove this, then rigid covariance, using the six essential potentials together with a rigid Euclidean coordinate system, will be locally equivalent to general covariance.

The uniqueness problem is related to identifying physical observers and this is related to finding the dynamical group of motion of General Relativity. In Newtonian Mechanics, this group is the group of rigid motions; that is, the group of transformations that depend on functions of one parameter, say $\lambda$, that leave the form $\bar{d}s^{2}=\bar{d}{\vec{x}}^{2}$ invariant. Beyond the rigid motions, the group we are looking for must leave the form $\bar{d}s^{2}= H^{-2}\bar{d}{\vec{x}}^{2}+\epsilon(\bar{d}\sigma)^{2}$ shape invariant (covariant). In \cite{Jaen2}, we studied  the case $\sigma=0$ and  $H\neq 0$, and we found that the group of motions was the homothetic group of motions. Surprisingly, as seen in \cite{Jaen2}, that group also plays a role in Newtonian Mechanics in relation to Newtonian cosmological questions. 

Now the problem that we face, leaving aside the cosmological potential, is that of finding the group of motions that leave $\bar{d}s^{2}=\bar{d}{\vec{x}}^{2}+\epsilon(\bar{d}\sigma)^{2}$ shape invariant. An important subgroup is the group of rigid motions. But now, in order to find the new required motions, we have no non-relativistic equivalent, as in the case of the homothetic group.

We hope that in the future we will be able to answer these questions.
%
%
%

\end{document}